\documentclass[aps,prl,twocolumn]{revtex4-1}

\usepackage{graphicx}
\usepackage{physics}
\usepackage{amssymb}
\usepackage{color}
\usepackage{ulem}

\begin{document}

\title{Universal Properties of Anisotropic Dipolar Bosons in Two Dimensions}

\author{J. S\'anchez-Baena$^{1,3}$, L. A. Pe\~na Ardila$^2$,
  G. E. Astrakharchik$^3$, F. Mazzanti$^3$}

\affiliation{$^1$ Center for Complex Quantum Systems, Department of
  Physics and Astronomy, Aarhus University, DK-8000 Aarhus C, Denmark}
\affiliation{$^2$ Institut für Theoretische Physik, Leibniz
  Universität Hannover, Germany} \affiliation{$^3$ Departament de
  F\'{\i}sica, Campus Nord B4-B5, Universitat Polit\`ecnica de
  Catalunya, E-08034 Barcelona, Spain}


\begin{abstract}

  The energy of ultra-dilute quantum many-body systems is known to
  exhibit a universal dependence on the gas parameter $x=n a_0^d$,
  with $n$ the density, $d$ the dimensionality of the space
  ($d=1,2,3$) and $a_0$ the $s$-wave scattering length.  The universal
  regime typically extends up to $x\approx 0.001$, while at larger
  values specific details of the interaction start to be relevant and
  different model potentials lead to different results. Dipolar
  systems are peculiar in this regard since the anisotropy of the
  interaction makes $a_0$ depend on the polarization angle $\alpha$,
  so different combinations of $n$ and $\alpha$
  value of the gas parameter $x$. In this work we analyze the scaling
  properties of dipolar bosons in two dimensions as a function of the
  density and polarization dependent scattering length up to very
  large gas parameter values. Using
  Quantum Monte Carlo (QMC) methods we study the energy and the main
  structural and coherence properties of the ground state of a gas of
  dipolar bosons by varying the density and scattering length for a
  fixed gas parameter. We find that the dipolar interaction shows
  relevant scaling laws up to unusually large values of $x$ that hold
  almost to the boundaries in the phase diagram where a transition to
  a stripe phase takes place.\\

\end{abstract}



\maketitle

Ultra-dilute systems have recently gained renewed interest since the
existence of liquid-like droplets of Bose mixtures was
predicted~\cite{Petrov_2015}, resulting in equilibrium densities
orders of magnitude lower than what is found in other systems such as
Helium~\cite{Barranco2006,Ancilotto2017}. In the context of Bose-Bose
mixtures, the formation of this liquid state results from the delicate
balance between the overall attractive mean-field energy arising
from the competition between interspecies attraction and
intraspecies repulsion, and a repulsive contribution caused by quantum 
fluctuations, which stabilizes the system. Bose-Bose self-bound droplets 
have been both described theoretically~\cite{Petrov_2015,Cikojevic_2018,
  Staudinger_2018} and observed experimentally~\cite{Cabrera_2018,
  Semeghini_2018}


  Ultradilute droplets have also been achieved in single-component
  dipolar systems. They result from the competition of the repulsion
  induced by a contact interaction term, and the dipole-dipole
  interaction (DDI)~\cite{Schmitt_2016, Bottcher_2019}. Dipolar droplets
  have also been predicted recently in dipolar
  mixtures~\cite{Bisset2021,Smith2021}. More complex systems featuring
  spin-orbit interactions have also been reported to be able to form
  ultralow density droplets that can even show a striped
  pattern~\cite{SanchezB_2020}.

In general, quantum systems at zero temperature
and very low densities are known to follow universal equations of
state~\cite{Lee_1957, Schick_1971, Giorgini_99, Mazzanti_2003}.  In
all cases, the leading terms are given by the mean-field (MF)
prediction, where the energy per particle $\epsilon(x)$ is a function
of the gas parameter $x = n a_0^d$, with $n$ the density, $d$ the
dimensionality of the space, and $a_0$ the $s$-wave scattering length
of the interatomic interaction. Within the lowest order Born
approximation, the only relevant parameter in a pseudo-potential
expansion of the interaction is $a_0$, while additional quantities
like the s-wave effective range or other parameters from higher order
partial waves do not contribute significantly. In the common case of
central interactions, the scaling in $x$ starts to break down as the
gas parameter exceeds a critical value $x_c \approx 0.001$. Below that
value, all interaction sharing the same $s$-wave scattering length
follow the MF + Beyond Mean Field prediction.



  Within mean-field theory, the energy per particle is
  linear with the density, $\epsilon=gn/2$, with the coefficient of
  proportionality given by the coupling constant $g$, which defines
  the strength of the short-range interaction between bosons modeled
  by a pseudopotential, $V_{SR}({\bf r})=g\delta({\bf r})$.  In three
  dimensions, the relation between the coupling constant and the
  $s$-wave scattering length is linear, $g_{3D}=4\pi\hbar^2a_0/m$,
  where $m$ is the particle mass, resulting in a linear dependence of
  $\epsilon$ on the gas parameter $x$, $\epsilon / (\hbar^2/ma_0^2) =
  2\pi x$.  The situation is significantly more complicated in the
  two-dimensional case
  where $\epsilon$ is fully defined by the density $n$, as
  experimentally shown for 2D Bose gases in
  Refs.~\cite{Hung2011,Dalibard2019}.


  
  In two dimensions, the dependence of the system
  properties on the $s$-wave scattering length $a_0$ occurs due to a
  quantum mechanical symmetry breaking. This is known as a quantum
  anomaly~\cite{Pitaevskii97,Olshanii2010},
  In the context of ultracold gases with short range interactions, this
  quantum anomaly manifests itself through the symmetry breaking of the scale
  invariance present in the classical field treatment of the problem.
  Therefore, the quantum anomaly phenomenon generates deviations from the
  predictions established from the classical field results. Among these, the
  modification of the frequency of the breathing mode in two dimensions has
  been both predicted theoretically for Bosons~\cite{Hofmann12, Yin20} and
  Fermions~\cite{Hu11,Olshanii2010} and observed experimentally in fermionic
  systems~\cite{Jochim2018, Vale2018}. Also, the change in the power-law
  exponents associated with long-range phase correlations in the system has
  been recently observed~\cite{Murthy19}. 
  Still the dependence of the coupling
  constant $g_{2D}$ on $a_0$ is extremely weak and comes through the
  logarithm of the gas parameter, $g_{2D}=4\hbar^2/(m|\ln
  x|)$~\cite{Schick71, Lieb2001}, with further perturbative terms
  introducing recursive contributions of the form $\ln |\ln
  x|$~\cite{Popov_1972, Lozovik78, Cherny2001, Mora2003,
    Pricoupenko2005}.
  A similar perturbative structure appears in the thermal effects
  associated with the Berezinskii-Kosterlitz-Thouless (BKT) phase
  transition~\cite{svistunov01,svistunov02}. However, in the latter
  case, the recursive term $\ln |\ln x|$ does not play a major role
  in typical experimental conditions, since its contribution is
  always smaller than that of a dimensionless experimental
  parameter~\cite{svistunov01}.  In fact, the weak dependence of the
  beyond mean field corrective terms limits the validity of the MF
  theory to exponentially small values of $x$.
  Indeed, it was shown in Ref.~\cite{Astra_2009} that it is necessary
  to reach values as small as $x \sim 10^{-100}$ to see that the influence
  of the beyond mean field corrections is in general negligible.
  As long as the energy is concerned, though, a cancellation of the
  logarithmic corrective terms leads to very small deviations from the
  MF prediction below $x\sim 10^{-3}$.
  The inclusion of dipolar physics brings a whole new degree of
  theoretical sophistication to a proper description of the equation
  of state $\epsilon$.  This is because, while technically speaking
  the dipolar interaction is short-ranged,
  its extension is large compared to other typical short-range potentials. 
On top of that, the dipole-dipole
interaction (DDI) depends not only on the distance but also on the
relative orientation of the constituents, introducing additional
degrees of freedom in the Hamiltonian.
In the specific case of polarized two-dimensional dipoles, the energy
per particle was shown in Ref.~\cite{Macia_2011} to follow the
universal prediction up to the critical value $x_c$.
Due to the anisotropy of the interaction, however, the $s$-wave
scattering length of polarized dipoles in 2D depends on the
polarization angle $\alpha$ as $a_0(\alpha)/a_d \approx
e^{2\gamma}(1-3\lambda^2/2)$, with $\lambda=\sin(\alpha)$,
$\gamma=0.577\ldots$ the Euler's constant, and $a_d=m C_{dd}/4\pi
\hbar^2$ the dipolar unit of length~\cite{Macia_2011}.  The DDI
potential
\begin{equation}
V_{dd}({\bf r}) = {C_{dd} \over 4\pi}
\left[ {1 - 3\lambda^2 \cos^2\theta \over r^3 } \right] \ ,
\label{Vdipdip_b1}
\end{equation}
describes dipoles moving in the XY plane
while being oriented externally by a polarization field 
contained in the XZ plane. It is important to
remark that increasing the tilting angle beyond a 
critical value $\alpha_c\simeq 0.615$ makes the system collapse.
This is because for $\alpha > \alpha_c$ 
the DDI becomes negative for values of $\theta$ around zero, 
while the quantum pressure is not enough to overcome it.

The dependence of $a_0$ on the polarization angle implies that the
same value of the gas parameter $x= na_0^2$ can be achieved in many
different ways by properly adjusting $n$ and $\alpha$. In particular,
increasing $\alpha$ leads to a reduction of the repulsion of the DDI,
thus implying that $n$ must be increased to keep $x$ constant. In
Ref.~\cite{Macia_2011} the authors showed that different combinations
leading to the same $x$ yield the same $\epsilon(x)$, even for low
values of $x$ that are larger than $x_c$. In this way, the equation of
state of bosonic dipoles in two dimensions seem to follow a universal
dipolar curve. Universal properties of dipolar bosons in two
dimensions have also been discussed in Ref.~\cite{Hofmann_2021}.

In this work we explore this universality among dipoles featuring
different orientations up to the large gas parameter values way above
$x_c$. We aim to characterize the degree of universality not only in
the energy per particle, but also in other observables directly
related to the structure of the system and its coherence properties.
This is done by performing diffusion Monte Carlo (DMC) simulations of
$N$ dipolar bosons moving in a box with periodic boundary conditions
contained in the XY plane. The system is described by the many-body
Hamiltonian
\begin{equation}
  H = -{\hbar^2 \over 2 m} \sum_{j=1}^N \nabla_j^2 +
  \sum_{i<j} V_{dd}({\bf r}_{ij}) \ ,
  \label{Hamiltonian_b1}
\end{equation}
with $m$ the mass and ${\bf r}_{ij}$ the relative position
vector. While DMC produces statistically exact energies, its
convergence properties benefit from the use of a variational guiding
wave function $\Phi_0({\bf r}_1, {\bf r}_2, \ldots, {\bf r}_N)$ to
drive the dynamics in imaginary time. In this work we build $\Psi_0$
as a Jastrow pair-product form
\begin{equation}
\Phi_0({\bf r}_1, {\bf r}_2, \ldots, {\bf r}_N) =
\prod_{i<j} f({\bf r}_{ij}) \ ,
\label{Jastrow_b1}
\end{equation}
with $f({\bf r})$ the solution of the zero-energy two body problem,
matched with a phononic tail at a distance $r_m$ that is variationally
optimized, as described in~\cite{Macia_2011}. As we wish to compare
different dipolar systems that have the same gas parameter $x$ and
different tilting angles, simulations are performed for different
values of the density $n$ as $\alpha$ changes such that $x$ remains
constant. We find that
$N=100$ particles are enough for all the gas
parameter values explored but $x=350$, where $N=200$ has been used. 
We have checked that our results remain essentially unchanged when
keeping $n$ constant while increasing $N$ and the box size
$L$.
We further elaborate on their dependency on the finite size of the system below.

\begin{figure}[t]
 \centering
 \includegraphics[width=1.00\linewidth]{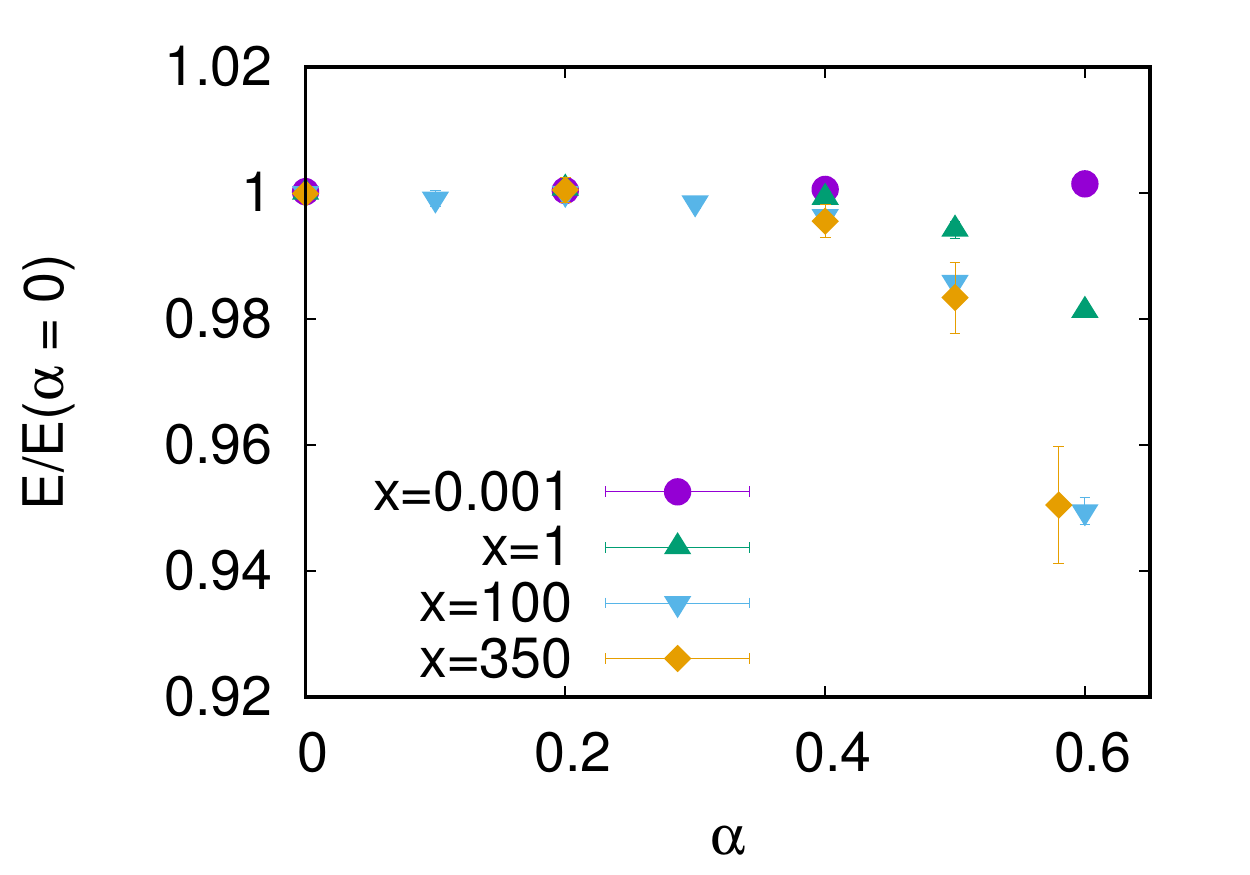}
 \caption{Ratio of the bulk
   DMC energy per particle, in units of the scattering
   length $a_0(\alpha)$ to $E(\alpha=0,x)$,
   for different values of the gas parameter $x$.
   The maximum tilting angle used
   for $x=350$ is $\alpha=0.58$. }
\label{energy_unversality_bulk}
\end{figure}

We start the discussion of the results by first addressing the energy
per particle of the system, since this is the driving quantity
characterizing universality among different quantum systems at zero
temperature.  In these systems, universality takes place when the
energy per particle, expressed in scattering length units, becomes a
function of the gas parameter $x=n a_0^2$ only. In the present case
where we compare the same (dipolar) interaction at different
polarization angles $\alpha$, universality implies that all ratios
$E(\alpha)/E_0(\alpha)$ with $E_0(\alpha)=\hbar^2/m a_0^2(\alpha)$,
must collapse to the same curve $\epsilon(x)$.
Figure~\ref{energy_unversality_bulk} shows the bulk DMC energies per
particle in dimensionless form for several values of the gas parameter
$x$ and polarization angle $\alpha$. We report
the ratio of the energy (in units of $E_0(\alpha)=\hbar^2/m
a_0^2(\alpha)$) to the energy at $\alpha=0$ (in units
$E_0(0)=\hbar^2/m a_0^2(0)$),
so that all curves in the figure 
start at one.  Notice that the maximum tilting angle explored for
$x=350$ is $\alpha=0.58$, as for larger values the ground state of the
system lays in the stripe phase~\cite{Macia_2014}. A perfect universal
behaviour would correspond to $E/E(\alpha=0) = 1$ for all polarization
angles where the system is still in the gas phase.  Surprisingly, and
as it can be observed from the figure, the universal behavior holds
for all $\alpha \lesssim 0.4$, while at larger angles slight
deviations less or equal than $5\%$ can only be seen at anomalously
large values of $x\gtrsim 100$, which lays orders of magnitude above
$x_c$.
In this sense, the energy for any value of $\alpha$ can be well
approximated with an error no larger than $5\%$ when its value at any
other single $\alpha$ (for instance $\alpha=0$) is known.
This scaling property allows for the computation of a single curve
that can be property rescaled and used as an input to alternative
mean-field models. We have found that a good fit to the DMC energies
is given by the expression
\begin{equation}
 \epsilon_{\text{full}} = E/N = \left( \frac{\hbar^2}{m a_0^2(\alpha)}
 \right) \exp \left[ A (\ln(x) + C)^l + B \right]
 \label{E_full}
\end{equation}
where $A=0.920, B=-7.917, C=8.0,$ and $l=1.117$ 
Furthermore and in agreement with 
Ref~\cite{Macia_2011}, in the universality regime of gas
parameter values $x \lesssim 10^{-3}$, the energy of the dipolar gas can
also be well approximated by the mean-field prediction
\begin{equation}
 \epsilon_{\text{MF}} = E_{\text{MF}}/N = \left( \frac{\hbar^2}{2 m
   a_0^2(\alpha)} \right) \frac{4 \pi x}{\abs{\ln(x)}} \ .
 \label{E_MF}
\end{equation}

\begin{figure}[t]
 \centering
\includegraphics[width=1\linewidth]{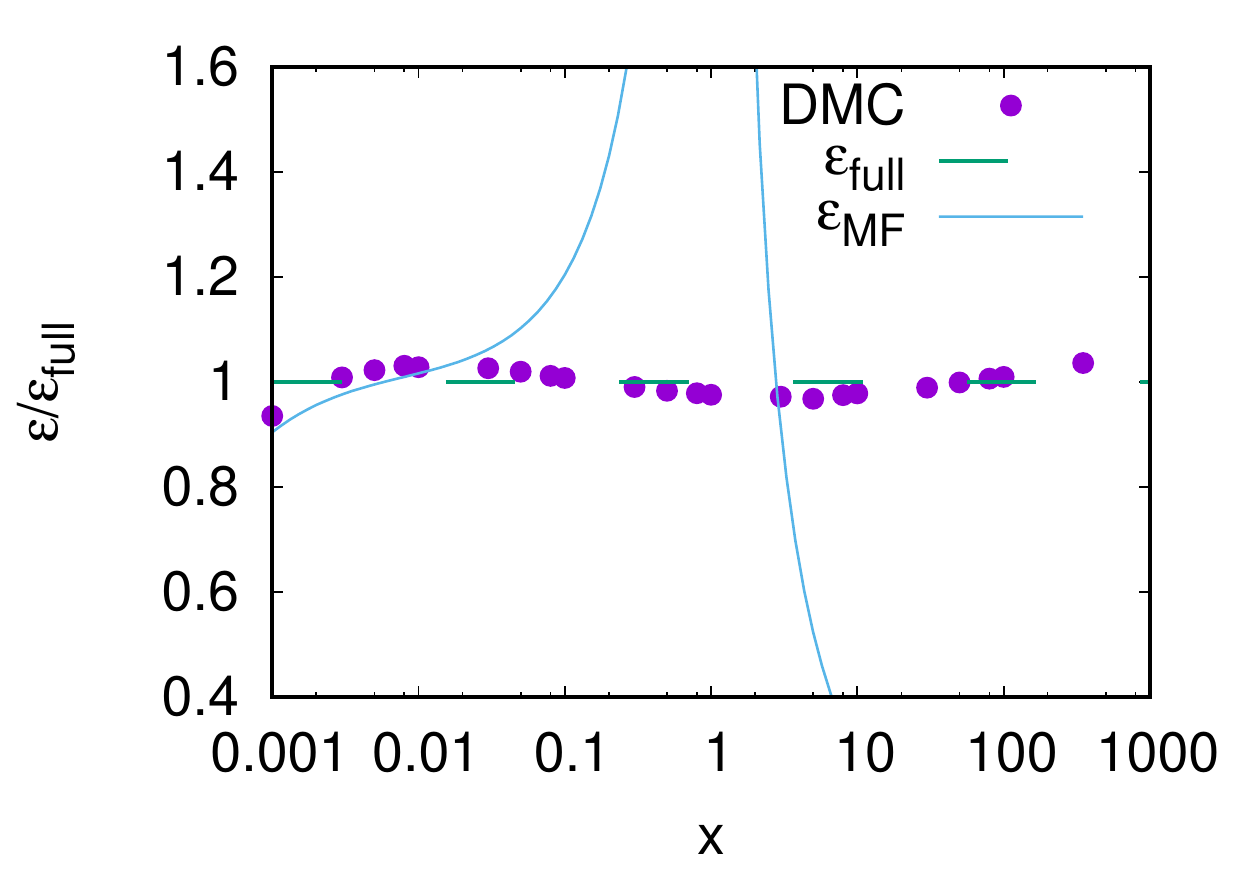}
\caption{ DMC energies per particle (symbols), 
    $\epsilon_{\text{full}}$ (dashed line) and $\epsilon_{\text{MF}}$ (solid line), 
    all in scattering length units, divided by $\epsilon_{\text{full}}$ as a function 
    of the gas parameter. Here, $\epsilon$ denotes an energy per particle.} 
\label{comparison_mf}
\end{figure}

We show in Fig.~(\ref{comparison_mf}) a comparison between the DMC
energies and the values obtained using the expressions in
Eqs.~(\ref{E_full}) and~(\ref{E_MF}). As it can be seen from the
Figure, the mean-field prediction of Eq.~(\ref{E_MF}) closely reproduces
the DMC energies for $x \lesssim 10^{-3}$, fairly close to the limit of
validity of the universal equation of state. Beyond that point, the
mean field functional drastically deviates from the DMC energies as
well as from the prediction of Eq.~(\ref{E_full}).


\begin{figure*}[t!]
 \centering
 \includegraphics[width=0.9\linewidth]{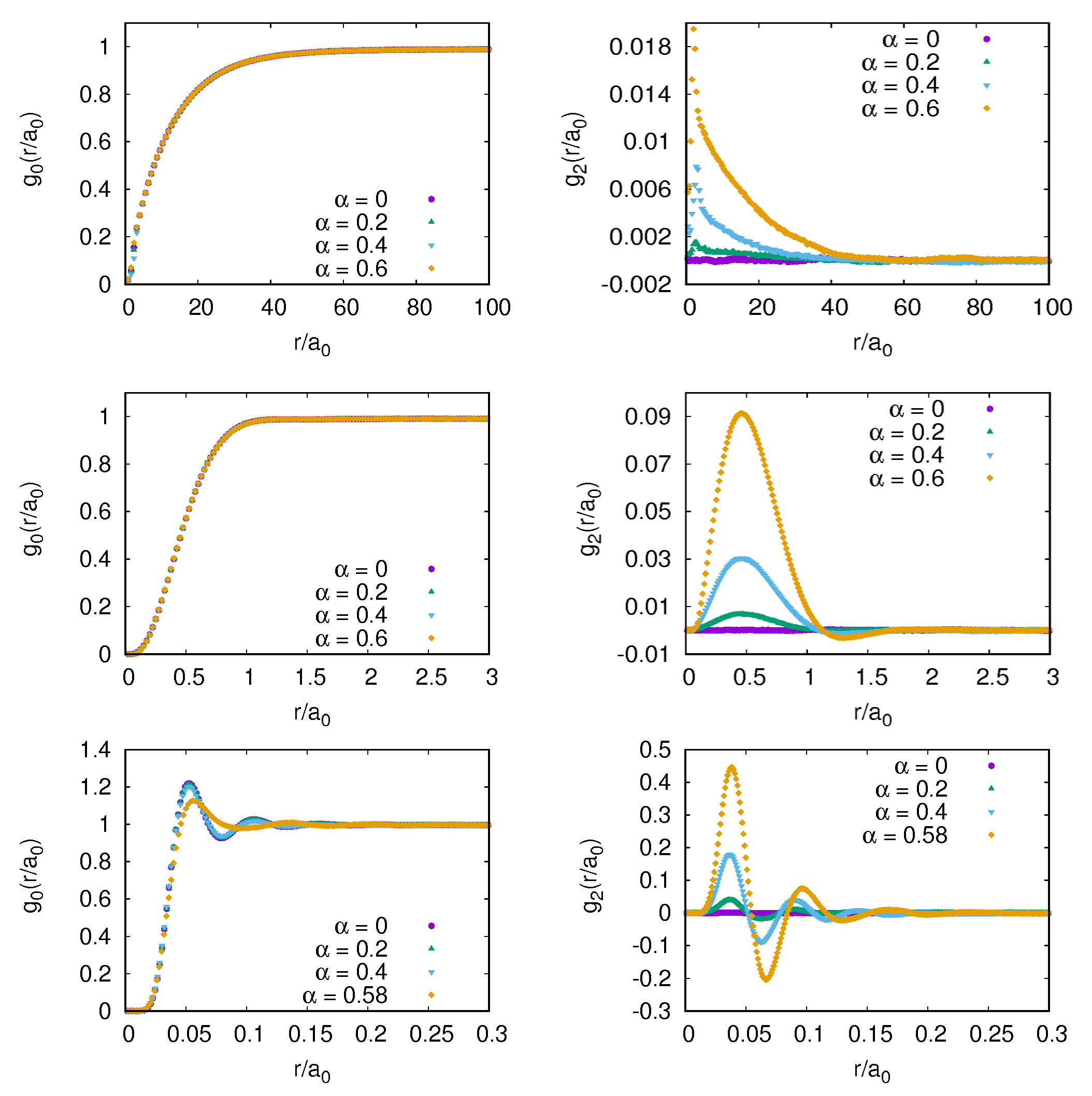}
 \caption{Isotropic (left plots) and first
     anisotropic modes (right plots) of the pair distribution function
     for $x=0.001$ (top), $x=1$ (middle) and $x=350$ (bottom) and for
     different values of the tilting angle.}
\label{gr_universality_bulk}
\end{figure*} 

It is also interesting to discuss the behavior of the energy expressed
in dipole units so that the energy scale is set to $E_0=\hbar^2/m
a_d^2$ for all polarization angles.  Since the scattering length
decreases with increasing $\alpha$ due to the anisotropy of the DDI,
one readily notices from the curves in
Fig.~\ref{energy_unversality_bulk} that the energy increases for
increasing $\alpha$, at least for $\alpha \lesssim 0.4$.  This may
seem to be a counteractive effect, as by
increasing $\alpha$ the interaction becomes less repulsive almost
everywhere. However, the density has to be increased
when the scattering length is reduced in order to keep the gas
parameter constant.  In this way, the net increase
in the energy is the result of two competing effects.

Next we discuss the structural properties of the system, starting with
the pair distribution function $g({\bf r})$, which is defined as
\begin{equation}
  g({\bf r}={\bf r}_{12}) = \frac{N(N-1)}{n N}
  {\int {\bf dr}_3 \cdots d{\bf r}_N\,
  \abs{\Psi({\bf r}_1, {\bf r}_2, \cdots, {\bf r}_N)}^2
  \over
  \int {\bf dr}_1 \cdots d{\bf r}_N\,
  \abs{\Psi({\bf r}_1, {\bf r}_2, \cdots, {\bf r}_N)}^2},
\end{equation}

This quantity measures the probability to find two particles at a
relative distance given by the position vector ${\bf r}$.  Considering
the anisotropy present in the system, it is convenient to perform a
partial waves expansion of $g({\bf r})$ in the form
\[
g({\bf r}) = \sum_{m=0}^\infty g_{2m}(r) \cos(2m \theta)
\]
with $(r, \theta)$ the polar coordinates. Due to the bosonic symmetry,
only even order modes contribute to this expansion.  In this way, the
emergence of anisotropic effects in the structure of the system is
manifested by the presence of non-vanishing 
$g_{2m}(r)$ terms with $m>1$. In practice, though, we have found that
higher order modes produce a negligible contribution when compared
with the first two.

We focus on two main aspects concerning the pair distribution
function: the effect of the anisotropy, and the possible scaling of
$g({\bf r})$ for different tilting angles $\alpha$. Results for
$g_0(r)$ and $g_2(r)$ are shown in the left and right panels of
Fig.~(\ref{gr_universality_bulk}) for increasing values of the gas
parameter and polarization angle.  In these plots all distances have
been scaled by the corresponding scattering lengths, which is
different for different values of $\alpha$.  As it can be seen from
the left upper and middle panels, for $x=0.001$ and $x=1$ the
isotropic modes are equal, regardless of the value of
$\alpha$. Similarly to the total energy discussed above, the pair
correlation functions follow a universal trend, even for values of the
gas parameter $x$ as large as 350, where deviations from a common
curve are evident only at the largest polarization angle considered,
$\alpha=0.58$. In this sense, the behavior of the isotropic mode of
the pair distribution function shows a universal dipolar behavior that
extends far beyond what is found in other quantum many-body
systems~\cite{Mazzanti_2003}.

The degree of anisotropy present can be measured by the strength of
the $g_2(r)$ mode, which is depicted in the right panels of the same
figure. As it can be seen, none of the curves are equal, not even at
the lowest value of $x=0.001$. This indicates that pure anisotropic
effects in $g({\bf r})$ do not scale, in contrast to what happens with
the isotropic mode. In any case, it should be noticed that the
relative strength of the anisotropic mode to the isotropic one is
always small in the range of $x$ and $\alpha$ values considered,
except for the largest ones. In this way one can conclude that the
impact on the anisotropy of the interaction in the spatial structure
of the system only affects significantly the dipolar gas close to the
transition to the stripe phase.

\begin{figure}[b]
 \centering
 \includegraphics[width=1.0\linewidth]{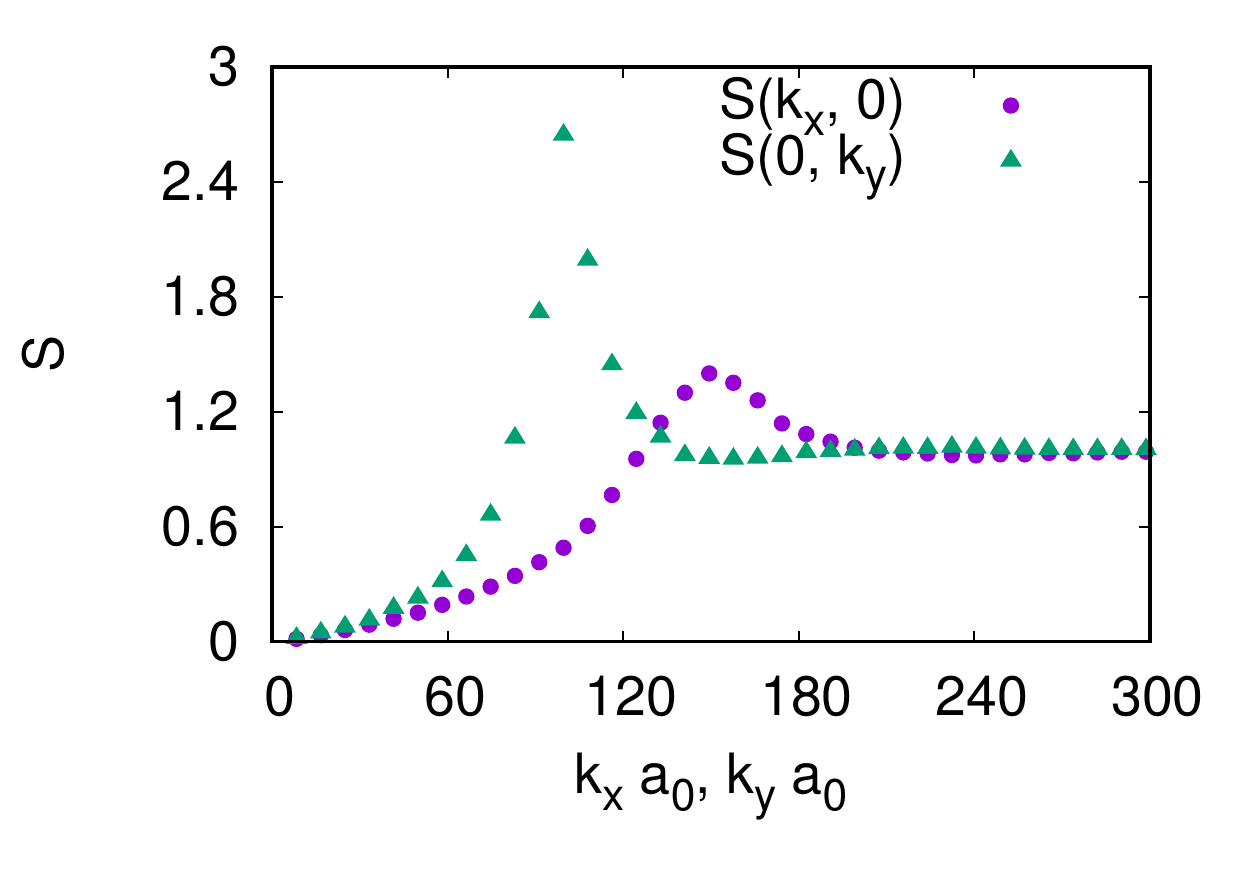}
 \caption{ Static Structure factor $S(k_x,0)$ and $S(0,k_y)$ computed for $x=350$,
   $\alpha=0.58$, with momenta scaled with the
   tilting-dependent scattering length $a_0(\alpha)$. }
 \label{sk_anisotropy_bulk}
\end{figure} 

From the pair distribution function one can obtain the static
structure factor $S({\bf k})$ by direct Fourier transform
\begin{equation}
S({\bf k}) = 1 + n \int d{\bf r} \,e^{i{\bf k}\,{\bf r}} \left( g({\bf r})-1\right) \ .
\label{Sk}
\end{equation}
This quantity characterizes spatial ordering in the system, as
periodic repetitions in space show up as peaks in $S({\bf k})$.  The
dipolar system is known to enter the stripe phase at large densities
and polarization angles. In this respect, the $x=350, \alpha=0.58$
point lays very close to the transition line~\cite{Macia_2014}.  Even
though this work is restricted to the study of dipolar gases,
the gas parameters and tilting angles explored reach values large
enough such that signs of spatial ordering along the direction of
maximal repulsion of the interaction are visible.  This is seen in
Fig.~(\ref{sk_anisotropy_bulk}), where we show $S(k_x,0)$ and
$S(0,k_y)$ for the largest values $x=350$, $\alpha=0.58$.  This
quantity has been obtained using the extrapolated estimator, which
corrects to first order the bias caused by the trial wave function in
the evaluation of expectation values of operators $\hat O$ that do not
commute with the Hamiltonian. The extrapolated estimator is computed
as the ratio $\langle \hat{O} \rangle_{\text{ext.}} = \langle \hat{O}
\rangle_{\text{DMC}}^2 /\langle \hat{O} \rangle_{\text{VMC}}$, where
the labels ``VMC" and ''DMC" stand for Variational and Diffusion Monte
Carlo, respectively.  As it can be seen,
$S(0,k_y)$, which corresponds to the direction of maximum repulsion,
shows a pronounced peak that is absent in $S(k_x,0)$.  This is the
triggering sign of spatial ordering along the $Y$ direction, in what
constitutes an anisotropic gas, a precursor of the supersolid stripe
phase. Being S({\bf k}) the Fourier transform of $g({\bf r})$, the
scaling properties presented by the static structure factor in terms
of $x$ and $\alpha$ are essentially the same ones presented by the
pair distribution function analyzed above.

\begin{figure}[t]
 \centering
 \includegraphics[width=1.0\linewidth]{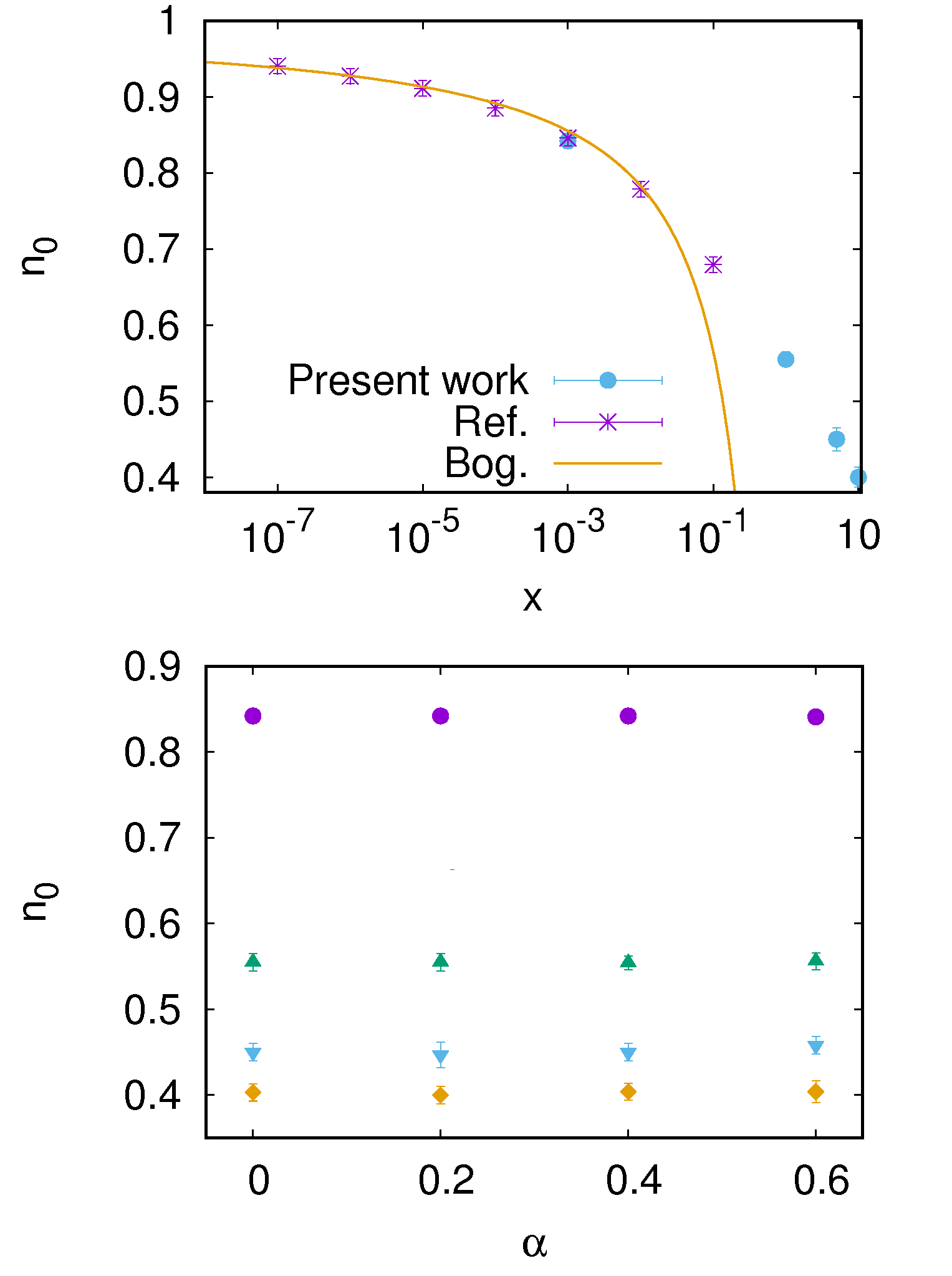}
 \caption{ Condensate fraction as a function of the gas parameter (upper plot),
   and as a function of $\alpha$ for different gas parameters 
   $x=0.001$ (dots), $1$ (up-triangles), $5$ (down-triangles), and $10$ (diamonds)
   (lower plot).
   In the upper plot, we provide the values obtained by Ref.~\cite{Macia_tesis}, as
   well as the Bogoliubov prediction, which is plotted for $x<1$.}
 \label{n0_universality_bulk}
\end{figure} 

In order to clarify the origin of the universality in the energy as a
function of the gas parameter, it is worth mentioning that the pair
distribution function can be directly related to the total energy per
particle of the system. In order to show that, we first notice that
the potential energy per particle can be written in the form
\begin{equation}
  \frac{ \langle V \rangle }{N} = \frac{n}{2} \int d{\bf r}\text{ }
  V_{dd}({\bf r}) g({\bf r}) \ .
\end{equation}
As shown in Fig.~\ref{gr_universality_bulk}, the anisotropic
contributions to the pair distribution function are much smaller than
the isotropic mode, so we can approximate $g({\bf r}) \simeq g_0({\bf r})$.
In dimensionless form, the potential energy per particle becomes

\begin{align}
 \frac{ \langle \tilde{V} \rangle }{N}  &= x \pi
 e^{-2 \gamma} \int d \tilde{r} \frac{g_0(\tilde{r})}{\tilde{r}^2} 
\end{align}
where $\tilde{r}$ and $\langle \tilde{V} \rangle$ are expressed in
units of $a_0(\alpha)$, $E_0(\alpha) = \hbar^2/(m a^2_0(\alpha))$,
respectively.
As a result, $\langle \tilde{V} \rangle/N$ depends only on the gas
parameter $x$ and on an integral of $g_0(\tilde{r}) =
g_0(\frac{r}{a_0(\alpha)})$, which is left almost unchanged for all
tilting angles. One thus concludes that, within this approximation,
universality in the pair distribution function induces universality in
the potential energy per particle. This is in agreement with the
universal properties of dipolar systems discussed in
Ref.~\cite{Hofmann_2021}. In order to link this result with the
universal behaviour of the total energy per particle, we use of the
Hellmann-Feynman theorem, which states that, for a Hamiltonian of the
form $\hat{H} = \hat{H}_0 + u \hat{H}_1$, the ground state energy can
be written as
\begin{equation}
 E = E_0 + \int_0^1 du\,\bra{\Psi(u)}\frac{d\hat{H}}{du}\ket{\Psi(u)} \ ,
\end{equation}
where $E_0$ is the ground state energy of $\hat{H}_0$. We now take
$\hat{H}_0$ and $\hat H_1$ to be the kinetic and (dipolar) potential
terms of the Hamiltonian, respectively.  The case $u=1$ corresponds to
the full Hamiltonian considered in this work. With this choice, $E_0 =
0$ as this corresponds to the ground state energy of a free gas of
bosons at zero temperature. In this way one has
\begin{align}
 {E \over N} & = \frac{1}{N} \int_0^1 du\text{ }\bra{\Psi(u)}V\ket{\Psi(u)} 
 \nonumber \\
 & = \frac{n}{2} \int_0^1 du\,
 \int d{\bf r}\, V_{dd}({\bf r}) g({\bf r},u) \ ,
\end{align}
or, in scattering length units
\begin{align}
  {\tilde{E} \over N} & = x \pi e^{-2 \gamma} \int_0^1 du 
  \int d\tilde{r} \frac{g_0(\tilde{r},u)}{\tilde{r}^2} \ .
 \label{HFT}
\end{align}
In this expression, $g({\bf r},u)$ stands for the pair distribution
function corresponding to a Hamiltonian $\hat{H} = \hat{H}_0 + u V$,
with $0<u<1$. Our previous analysis has shown that already for $u=1$
the contribution of the isotropic term dominates the pair distribution
function in the range of tilting angles and gas parameters considered.  By
reducing the strength of the dipolar interaction with $u<1$, as given
in Eq.~(\ref{HFT}), the impact of the anisotropic modes is
reduced as the potential contribution to the total energy is less
relevant the lower $u$ is.
Therefore, Eq.~(\ref{HFT}) links the universality of the pair
distribution function to the universality of the total energy per
particle. It must be remarked, however, that the low contribution of the
anisotropic modes of the pair distribution function, compared to the
isotropic one, is key to ensure universality in the energy, since the
anisotropic modes are not universal as seen in
Fig.~\ref{gr_universality_bulk}. We conclude that the
lack of strong anisotropic contributions in the structure of the
system, even at large gas parameters and tilting angles, leads to
universality in the total energy per particle. We further extend over
this argument below.

Next, we discuss the condensate fraction $n_0$ of the system. 
This quantity is obtained from the large-distance limit of the 
off-diagonal one body density matrix
\begin{equation}
\!\!\!\!\!\!\!
\rho_1({\bf r}) = 
N {\int d{\bf r}_2 \cdots d{\bf r}_N 
\Psi^*({\bf r}_1 + {\bf r}, {\bf r}_2, \cdots, {\bf r}_N)
\Psi({\bf r}_1, {\bf r}_2, \cdots, {\bf r}_N)
\over
\int d{\bf r}_1 \cdots d{\bf r}_N 
\mid\! \Psi({\bf r}_1, \cdots, {\bf r}_N) \!\mid^2
}
\label{rho1}
\end{equation}
as $n_0=\rho_1(|{\bf r}|\to\infty)/n$.  The upper plot of
Fig.~(\ref{n0_universality_bulk}) shows $n_0$ for different values of
the gas parameter $x$ and tilting angle $\alpha$. Remarkably, the
condensate fraction remains essentially constant at fixed $x$.  We
find an almost perfect scaling behaviour up to the largest value
$x=10$ explored, which stays largely away from the diluteness regime.
As expected, the value of $n_0$ decreases with increasing $x$, as the
enhancement of quantum fluctuations at larger densities favors the
depletion of the condensate.  In order to discern whether the
dependence of the condensate fraction of $x$ is universal or not, we
also to compare our results to the Bogoliubov prediction
$n_0^{\text{B}} = 1 - 1/\abs{\text{ln} x}$.
As it can be seen, the Bogoliubov prediction is recovered only in the
weakly interacting regime corresponding to $x\lesssim 0.001$, while
significant deviations appear as $x$ increases. Still and as mentioned
above, the condensate fraction has the same value for fixed $x$ and
different $\alpha$, thus showing a clear scaling behavior as the
previous quantities analyzed. For large $x$ the DMC prediction is
significantly larger than the values obtained in the Bogoliubov model,
as expected.

\begin{figure}[t]
 \centering
 \includegraphics[width=1.0\linewidth]{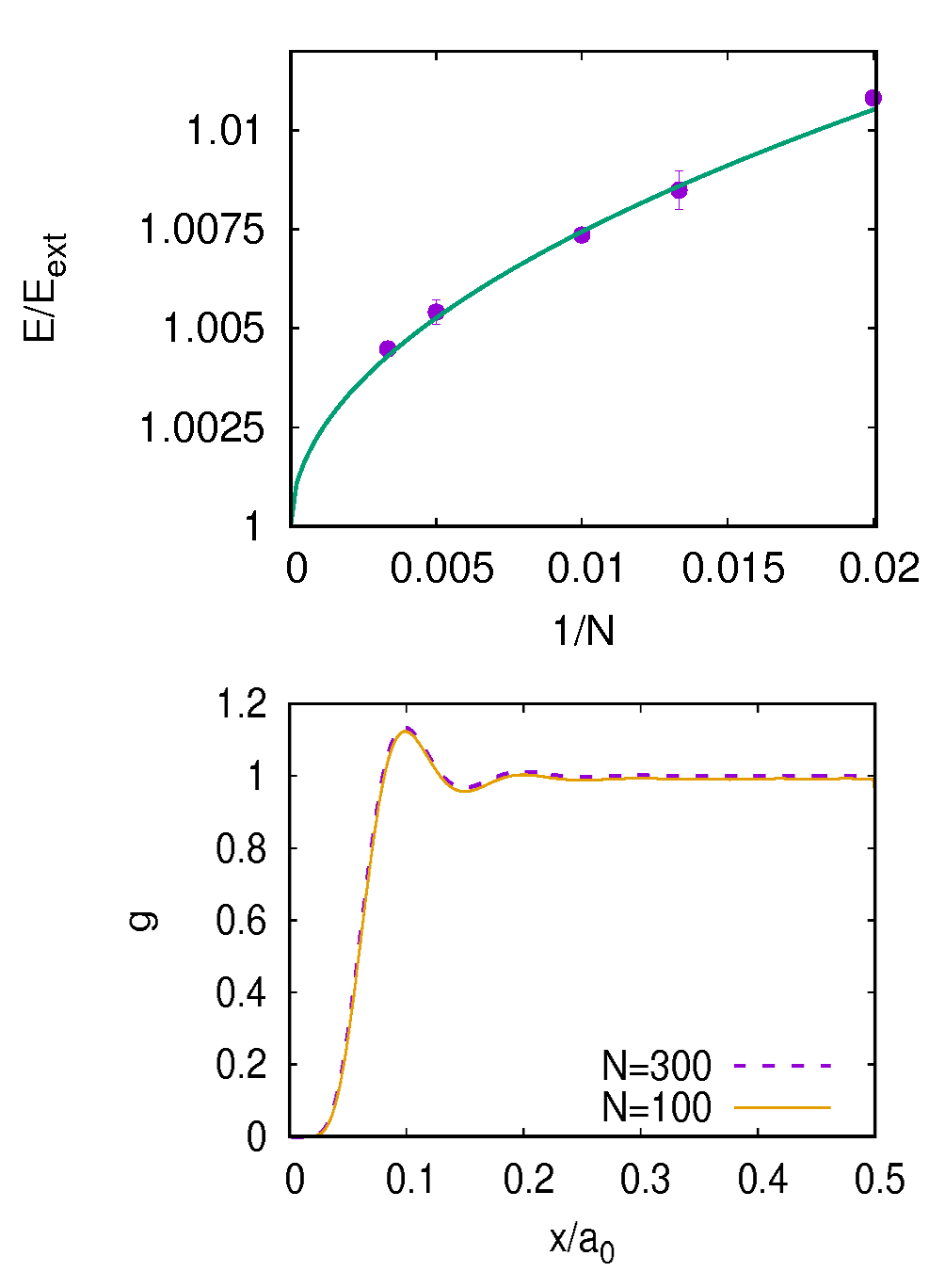}
 \caption{
   (Top) ratio of the DMC energies to the extrapolated $N\to\infty$
   value for $x=100$ and $\alpha=0$.  The solid line corresponds to a
   fit of the form $f(N) = E_{\text{ext}} + C/\sqrt{N}$ from where the
   extrapolated value is obtained. (Bottom) Pair distribution function
   at $x=100$ and $\alpha=0$ for $N=100$ and $N=300$ particles at the
   same density. }
 \label{size_effect}\end{figure} 

\begin{figure}[b]
 \centering
 \includegraphics[width=0.9\linewidth]{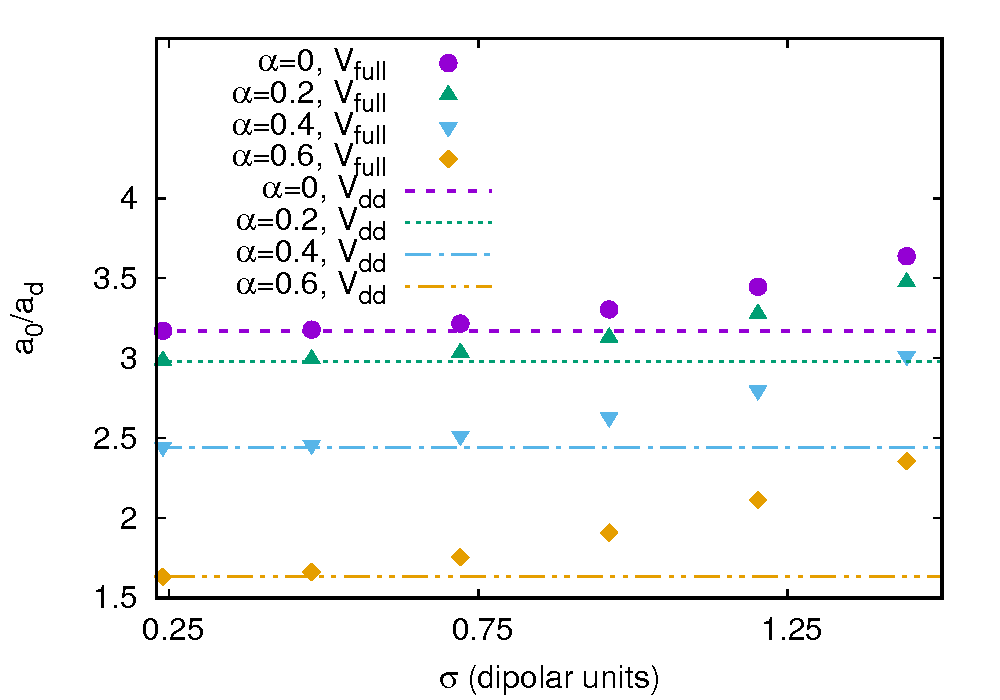}
 \caption{Scattering length of the compound system described by the
   Hamiltonian in Eq.~(\ref{Hamiltonian_b1_sr}) corresponding to a
   dipolar system with an additional Van der Waals tail, as a function
   of $\sigma$ and for different polarization angles. Here, $\sigma$
   is given in dipolar units, with the characteristic length and
   energy scales given by $l = a_d$, $\epsilon_l = \hbar^2/m a_d^2$
   respectively.}
 \label{Scatt_Length_fullH}
\end{figure} 

\begin{figure*}[t]
 \centering
 \includegraphics[width=0.9\linewidth]{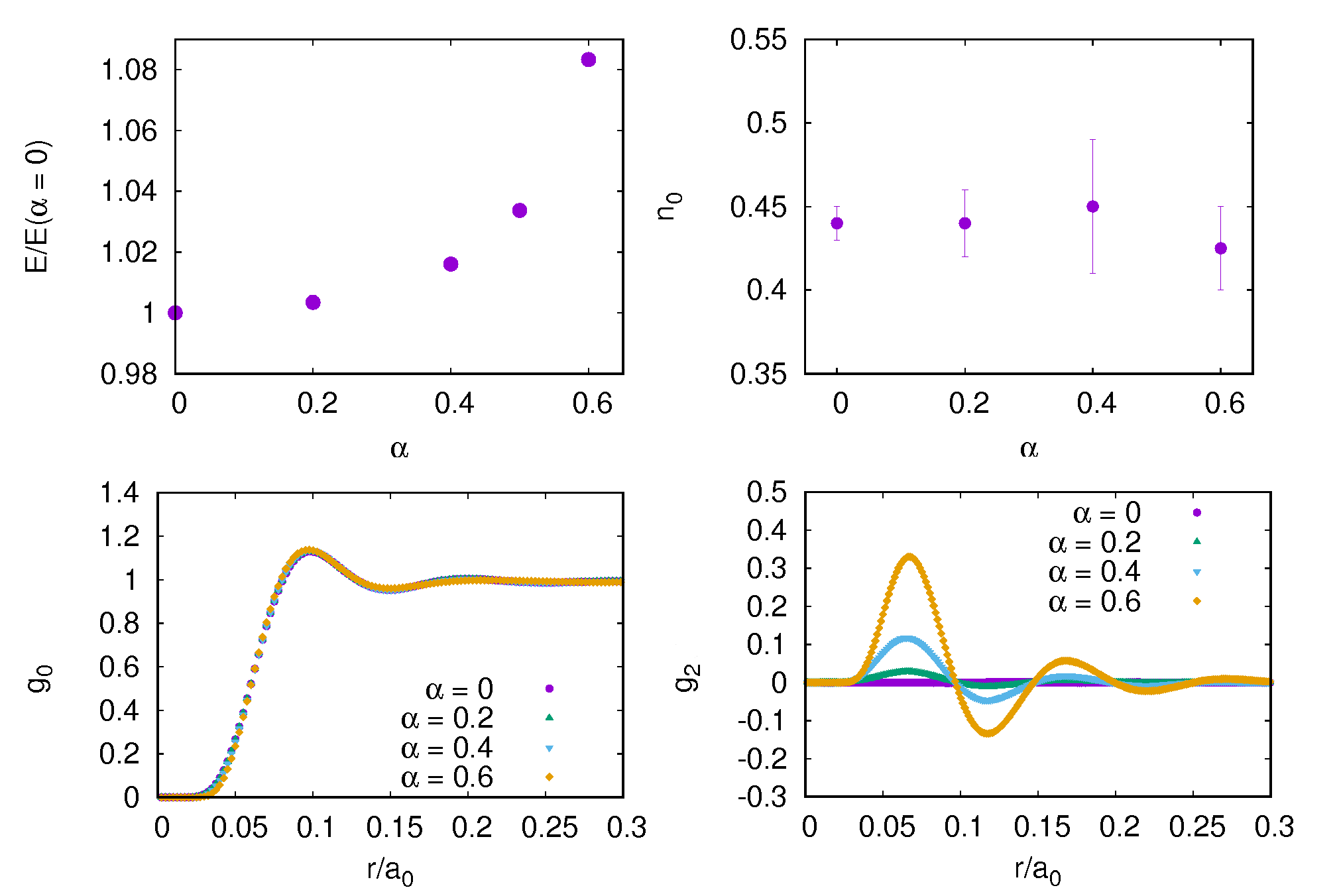}
 \caption{ Ratio of the bulk DMC energy per
     particle, in units of the scattering length $a_0(\alpha)$ to
     $E(\alpha=0,x)$, for $x=100$ (top left panel). Condensate
     fraction as a function of the tilting values $\alpha$ for $x=5$
     (top right panel). Isotropic (bottom left panel) and first
     anisotropic modes (bottom right panel) of the pair distribution
     function for $x=100$. All quantities have been computed for the
     Hamiltonian of Eq.~\ref{Hamiltonian_b1_sr}, where a short range
     repulsive potential is considered along the DDI. }
 \label{universality_sr}
\end{figure*}

  To conclude the numerical analysis, and
  considering all simulations have been performed with a fixed number
  of particles, we analyze how finite size effects
  influence our results.
  We report in the top panel of Fig.~\ref{size_effect} the ratio of
  the energy of the dipolar gas at fixed $N$ to the corresponding
  value extrapolated to the thermodynamic limit ($N\to \infty$),
  denoted by $E_{\text{ext}}$, for $x=100$ and $\alpha=0$.  We also
  show the fit from where the extrapolated value is obtained, which
  corresponds to a function of the type $f(N) = E_{\text{ext}} +
  C/\sqrt{N}$. This is consistent with finite size calculations
  performed in Ref.~\cite{Astrakharchik_07} for a system of non-tilted
  dipoles in two dimensions. Larger tilting angles produce similar
  results.  In much the same way and in order to characterize the
  influence of finite size effects in other static properties, we
  report on the bottom panel of the same figure the pair distribution
  function for $N=100$ and $N=300$, also at $x=100$ and $\alpha=0$.
  In all cases, the box size $L$ is chosen such that the density is
  kept constant while changing $N$.
  As can be seen, finite size corrections for both quantities are
  small and below $0.75\%$ already at $N=100$, thus confirming that the
  universality conditions described above hold also in the
  thermodynamic limit.

After analyzing universality in the energetic and structural properties
  of the purely dipolar system, it is interesting to discuss what the
  effect of adding a short range isotropic interaction is. This is a
  relevant issue, as in actual experiments on dipolar systems
  this term is usually present.
  When it is included, the Hamiltonian becomes
\begin{align}
  H &= -{\hbar^2 \over 2 m} \sum_{j=1}^N \nabla_j^2 +
  \sum_{i<j} V_{dd}({\bf r}_{ij}) + V_{sr}({\bf r}_{ij}) \nonumber \\
  &= -{\hbar^2 \over 2 m} \sum_{j=1}^N \nabla_j^2 +
  \sum_{i<j} V_{full}({\bf r}_{ij}) \ ,
  \label{Hamiltonian_b1_sr} 
\end{align}
  where $V_{full}({\bf r})$ stands for the total potential acting on
  the atoms.  In order to explore how the inclusion of this short
  range term affects universality, we have chosen a model interaction
  of the form $V_{sr}(r_{ij}) = \left( \sigma/r_{ij} \right)^6$.  At
  this point, we have calculated the scattering length of the
  Hamiltonian in Eq.~(\ref{Hamiltonian_b1_sr}) from the large distance
  asymptotic behavior of the zero-energy solution of the two-body
  problem as discussed in~\cite{Khuri_2009}. This treatment is
  equivalent to solving the Scattering T-matrix of the system, and
  taking its zero-momentum limit. The result is shown in
  Fig.~\ref{Scatt_Length_fullH}. As it can be seen, although non-zero,
  the addition of the short-range term, for moderate values of $\sigma$, does not alter significantly
  the scattering properties with
  respect to the purely dipolar model.  In
  order to confirm that in the many-body case, we have performed additional
  simulations corresponding to the Hamiltonian in
  Eq.~(\ref{Hamiltonian_b1_sr}) with the chosen $V_{sr}(r)$. We have
  set $\sigma = 0.25$ in dipolar units, as in this case the presence of $V_{sr}({\bf
    r})$ does not modify appreciably the total scattering length.  
  We have computed the energy per particle
  and the pair distribution function at $x=100$, and the condensate
  fraction at $x=5$.
  The results are shown in Fig.~\ref{universality_sr}, where we
  report, for $x=100$ and different values of the tilting angle, the ratio
  $E/E(\alpha=0)$ (as in Fig.~\ref{energy_unversality_bulk}), 
  the $g_0(r)$ and $g_2(r)$ modes of the pair
  distribution function (as in Fig~\ref{gr_universality_bulk}),
  and the condensate fraction (as in the lower panel of
  Fig~\ref{n0_universality_bulk}) for $x=5$. As we can see from the
  figure, universality still holds in the main static properties of the
  system.
  The energy per particle departs only a few percent from the perfect
  universal behaviour (corresponding to $E/E(\alpha=0) = 1$), and only
  at the largest polarization becomes slightly noticeable.  The pair
  distribution function keeps being dominated by the isotropic mode,
  which clearly shows universality. In much the same way, the
  condensate fraction does not present a noticeable dependence on the
  tilting angle.  We thus conclude that the presence of a short range
  repulsive potential does not alter significantly the universality
  present in the 2D dipolar gas when the DDI interaction is considered
  alone.

In view of the results reported in this work, we
  conclude that the universality displayed by the 2D dipolar gas is a
  direct consequence of the fact that the gas parameter must be
  increased to very large values ($x \simeq 400$) for the anisotropy
  of the DDI to have a strong influence in the structural properties
  of the system. This is furtherly supported by the fact that one can
  recover the properties reported in this work to a high degree of
  accuracy by replacing the full DDI by an isotropic potential of the
  form
  \begin{equation}
    V_{eff}(r, \alpha) = \frac{C_{dd}}{ 4 \pi } \frac{ \left( 1
      - \frac{3}{2} \sin^2 \alpha \right) }{ r^3 } \ ,
    \label{Vsr_r}
  \end{equation}
  where the tilting angle becomes simply a parameter that tunes the
  effective strength of the interaction. This particular form of the
  potential makes its scattering length be almost identical to that of
  the fully anisotropic DDI interaction~\cite{Macia_2011}.  When the
  many-body Schr\"odinger equation for $V_{eff}(r)$ is expressed in
  scattering length units, it becomes independent of the polarization
  angle $\alpha$
  \begin{equation}
    -\frac{1}{2} \sum_{i=1}^N \tilde{\nabla}_i^2 \Psi + \sum_{i<j}
    \frac{ e^{-2 \gamma} }{ \abs{{\bf \tilde{r}}_{ij}}^3 } \Psi = \tilde{E} \Psi \ ,
    \label{schrodinger_eff_units}
  \end{equation}
  meaning that, in these units, solving
  Eq.~(\ref{schrodinger_eff_units}) yields $\alpha$-independent
  results.  Furthermore, since $V_{eff}(r, \alpha=0) = V_{dd}(r,
  \alpha=0)$, the properties obtained when solving
  Eq.~(\ref{schrodinger_eff_units}) are the same as those obtained when
  solving the Schr\"odinger equation for the full $V_{dd}(r,
  \alpha=0)$, written in scattering length units. Thus, one can see
  from the results in
  Figs.~\ref{energy_unversality_bulk},~\ref{gr_universality_bulk}
  and~\ref{n0_universality_bulk} that the energy per particle, the
  radial distribution function, and the condensate fraction of the
  system obtained when the potential $V_{eff}(r, \alpha)$ is
  considered (which correspond to the data for $\alpha=0$ in the
  Figures) approximate reasonably well the results obtained with the
  full DDI when $\alpha \neq 0$. We believe that the fact that this
  isotropic approximation is successful, which is a consequence of
  universality, is due to the lack of a strong anisotropic influence 
  in the structural properties of the system.

To summarize, we have studied the scaling of the dipolar interaction
as a function of the polarization angle $\alpha$ and gas parameter $x$
in a system of two-dimensional bosonic dipoles.  We have found that
universality is lost already at $x\approx 0.001$ where the energy per
particle deviates from the mean-field prediction as expected. Beyond
that point, however, all energy curves collapse to a single one when
properly scaled by the tilting-dependent scattering length
$a_0(\alpha)$. This behavior holds up to surprisingly large values of
$x$ close to the gas-stripe transition line, like $x=350$, and up to
large polarization angles near the collapse limit. In this same
region, this scaling property is not only present in the energy, but
also on the condensate fraction for all polarization angles considered
($\alpha \in [0,\text{ }0.6]$), and in the most relevant structural
properties like the pair distribution function and the static
structure factor.  We have also shown that this
  behaviour is still present in the system when a short range
  repulsive potential is considered along with the DDI, which is
  typically the case in actual experiments.  All this means that, for
any $\alpha$ contained in the region considered, the angular
dependence (and thus the anisotropic features) of the most relevant
static properties of dipolar quantum Bose gases in two dimensions are
entirely contained in the $\alpha$-dependent $s$-wave scattering
length, which is well approximated by the expression $a_0(\alpha)/a_d
\approx e^{2\gamma}(1-3\lambda^2/2)$ with $\lambda=\sin(\alpha)$ and
$a_d$ setting the dipolar length scale.
From our analysis we finally conclude that the
  universal behavior of the dipolar Bose gas in two dimensions is a
  consequence of the overall low impact of the anisotropy on the
  structural properties of the system up to astonishingly large values
  of the gas parameter.


\begin{acknowledgments}
  {\em Acknowledgements:}
The work has been supported by grant PID2020-113565GB-C21 from
  MCIN/AEI/10.13039/501100011033,
and by the  Danish  National
Research  Foundation  through  the  Center  of  Excellence
“CCQ” (Grant agreement no.:  DNRF156). J. S\'anchez-Baena acknowledges funding by the European Union, the Spanish Ministry of Universities and the Recovery, Transformation and Resilience Plan through a grant from Universitat Politecnica de Catalunya.
\end{acknowledgments}


\begin{references}

\bibitem{Petrov_2015}
  D. S. Petrov,
  Phys. Rev. Lett. {\bf 115}, 155302 (2015).

\bibitem{Barranco2006}
  M. Barranco, R. Guardiola, E. S. Hernández, R. Mayol, J. Navarro, and M. Pi, J.
  Low Temp. Phys. {\bf 142}, 1 (2006). 

\bibitem{Ancilotto2017}
  F. Ancilotto, M. Barranco, F. Coppens, J. Eloranta, N. Halberstadt, A. Hernando,
  D. Mateo, and M. Pi.,
  Int. Rev. Phys. Chem. {\bf 36}, 621 (2017)

\bibitem{Cikojevic_2018}
  V. Cikojevi\'c, K. D\v{z}elalija, P. Stipanovi\'c, L. V. Marki\'c, and
  J. Boronat,
  Phys. Rev. {\bf B97}, 140502(R) (2018).

\bibitem{Staudinger_2018}
  C. Staudinger, F. Mazzanti, and R. E. Zillich,
  Phys. Rev. {\bf A98}, 023633, (2018).

\bibitem{Cabrera_2018}
  C. R. Cabrera, L. Tanzi, J. Sanz, B. Naylor, P. Thomas, P. Cheiney,
  and L. Tarruell,
  Science {\bf 359}, 301–304 (2018).

\bibitem{Semeghini_2018}
  G. Semeghini, G. Ferioli, L. Masi, C. Mazzinghi, L. Wolswijk,
  F. Minardi, M. Modugno,G. Modugno, M. Inguscio, and M. Fattori,
  Phys. Rev. Lett. {\bf 120}, 235301 (2018).


\bibitem{Schmitt_2016}
  M. Schmitt, M. Wenzel, F. Böttcher, I. Ferrier-barbut, and T. Pfau,
  Nature {\bf 539}, 259 (2016).

\bibitem{Bottcher_2019}
  F. B\"ottcher, M. Wenzel, J-N. Schmidt, M. Guo, T. Langen,
  I. Ferrier-Barbut, T. Pfau, R. Bombín, J. S\'anchez-Baena,
  J. Boronat, and F. Mazzanti,
  Phys. Rev. Research {\bf 1}, 033088 (2019).

\bibitem{Bisset2021}
  R. R. Bisset, L. A. Pe\~na Ardila, and L. Santos,
  Phys. Rev. Lett. {\bf 126} 025301 (2021).

\bibitem{Smith2021}
  J. C. Smith, D. Baillie, and P. B. Blakie,
  Phys. Rev. Lett. {\bf 126} 025302 (2021)
  
\bibitem{SanchezB_2020}
  J. S\'anchez-Baena, J. Boronat, and F. Mazzanti,
  Phys. Rev. {\bf A102}, 053308 (2020).

\bibitem{Lee_1957}
  T.D. Lee and C.N. Yang,
  Phys. Rev. {\bf 105}, 1119 (1957)
  T.D. Lee, K. Huang, and C.N. Yang, {\em ibid.} {\bf 106}, 1135
  (1957).

\bibitem{Schick_1971}
  M. Schick, Phys. Rev. A {\bf 3}, 1067 (1971);
  E. H. Lieb, and J. Yngvason, J. Stat. Phys. {\bf 103}, 509 (2001).
  
\bibitem{Giorgini_99}
  S. Giorgini, J. Boronat, and J. Casulleras,
  Phys. Rev. {\bf A60}, 5129 (1999).

\bibitem{Mazzanti_2003}
  F. Mazzanti, A. Polls, and A. Fabrocini, 
  Phys. Rev. {\bf A67}, 063615, (2003).
  
\bibitem{Hung2011}
  C.-L. Hung, X. Zhang, N. Gemelke, and C. Chin,
  Nature  \textbf{470}, 236 (2011).  

\bibitem{Dalibard2019}
  R. Saint-Jalm, P. C. M. Castilho, \'E. Le Cerf,
  B. Bakkali-Hassani, J.-L. Ville, S. Nascimbene, J. Beugnon, and J. Dalibard,
  Phys. Rev. X \textbf{9}, 021035 (2019)
 
\bibitem{Pitaevskii97}
  L. P. Pitaevskii and A. Rosch,
  Phys. Rev. A \textbf{55}, R853 (1997).

\bibitem{Olshanii2010}
  M. Olshanii, H. Perrin, and V. Lorent,
  Phys. Rev. Lett. \textbf{105}, 095302

\bibitem{Hofmann12}
J. Hofmann
Phys. Rev. Lett. \textbf{108}, 185303 (2012)

\bibitem{Yin20}
X. Y. Yin, H. Hu, and X. Liu
Phys. Rev. Lett. \textbf{124}, 013401 (2020)

\bibitem{Hu11}
Y. Hu and Z. Liang
Phys. Rev. Lett. \textbf{107}, 110401 (2011)
 
\bibitem{Jochim2018}
  M. Holten, L. Bayha, A. C. Klein, P. A. Murthy, P. M. Preiss, and S. Jochim,
  Phys. Rev. Lett. \textbf{121}, 120401 (2018).

\bibitem{Vale2018}
  T. Peppler, P. Dyke, M. Zamorano, I. Herrera, S. Hoinka, and C. J. Vale,
  Phys. Rev. Lett. \textbf{121}, 120402 (2018)
  
\bibitem{Murthy19}
  P. Murthy, N. Defenu, L. Bayha, M. Holten, P. Preiss, T. Enss, S. Jochim,
  Science \textbf{365}, 268-272 (2019)
 
\bibitem{Schick71} M. Schick,
  Phys. Rev. A \textbf{3}, 1067 (1971).

\bibitem{Lieb2001}
  E. H. Lieb, R. Seiringer, and J. Yngvason,
  Commun. Math. Phys. \textbf{224}, 17 (2001)

\bibitem{Popov_1972}
  V. N. Popov, Theor. Math. Phys. {\bf 11}, 565 (1972);
  V. N. Popov, {\em Functional Integrals in Quantum Field Theory and
    Statistical Physics} (Reidel, Dordrecht, 1983).

\bibitem{Lozovik78}
  Yu. E. Lozovik and V. I. Yudson,
  Physica A \textbf{93}, 493 (1978).

\bibitem{Cherny2001}
  A. Y. Cherny and A. A. Shanenko,
  Phys. Rev. E \textbf{64}, 027105

\bibitem{Mora2003}
  C. Mora and Y. Castin,
  Phys. Rev. A \textbf{67}, 053615 (2003).

\bibitem{Pricoupenko2005}
  L. Pricoupenko,
  Phys. Rev. A \textbf{70}, 013601 (2004).

\bibitem{svistunov01}
  N. Prokof'ev, O. Ruebenacker, and B. Svistunov,
  Phys. Rev. Lett. \textbf{87}, 270402 (2001).

\bibitem{svistunov02}
  N. Prokof’ev and B. Svistunov,
  Phys. Rev. A \textbf{66}, 043608 (2002).

\bibitem{Astra_2009}
  G. E. Astrakharchik, J. Boronat, J. Casulleras, I. L. Kurbakov, and Yu. E. Lozovik
  Phys. Rev. A{\bf 79}, 051602(R) (2009).

\bibitem{Petrov01}
  D. S. Petrov and G. V. Shlyapnikov,
  Phys. Rev. A \textbf{64}, 012706 (2001)

\bibitem{Ville18}
  J. L. Ville, R. Saint-Jalm, \'E. Le Cerf, M. Aidelsburger,
  S. Nascimb\`ene, J. Dalibard, and J. Beugnon,
  Phys. Rev. Lett. \textbf{121}, 145301 (2018)

\bibitem{Mazzanti10}
  G. E. Astrakharchik, J. Boronat, I. L. Kurbakov, Yu. E. Lozovik, and
  F. Mazzanti,
  Phys. Rev. A \textbf{81}, 013612 (2010)

\bibitem{Macia_2011}
  A. Macia, F. Mazzanti, J. Boronat, and R. E. Zillich, 
  Phys. Rev. {\bf A84}, 033625 (2011).
  
\bibitem{Hofmann_2021}
  J. Hofmann, and W. Zwerger
  Phys. Rev. Research {\bf 3}013088 (2021).  


\bibitem{Astrakharchik_07}
 G. E. Astrakharchik, J. Boronat, I. L. Kurbakov, and Yu. E. Lozovik
Phys. Rev. Lett. \textbf{98}, 060405 (2007)
  
\bibitem{Macia_2014}
  A. Macia, F. Mazzanti and J. Boronat, 
  Phys. Rev. {\bf A90}, 061601(R) (2014).  
  
\bibitem{Macia_tesis}
  A. Macia,
  \textit{Microscopic description of two dimensional dipolar
    quantum gases}, Barcelona, Academic (2015).

\bibitem{Khuri_2009}
  N. N. Khuri, A. Martin, J.-M. Richard, and Tai Tsun Wu
  J. Math. Phys. {\bf 50} 072105 (2009).
\end{references}

\end{document}